\documentclass[prd,twocolumn,nofootinbib,superscriptaddress]{revtex4-2}

\usepackage{preamble}

\begin{document}

\title{Gravitational wave signal and noise response of an optically levitated sensor in a Fabry–Pérot cavity}

\author{Andrew Laeuger\,\orcidlink{0000-0002-8212-6496}}
\email{alaeuger@caltech.edu}
\affiliation{Department of Physics and Astronomy, Caltech, Pasadena, California 91125, USA}

\author{Shafaq Gulzar Elahi}
\affiliation{Center for Fundamental Physics, Department of Physics and Astronomy,
Northwestern University, Evanston, IL 60208, USA}
\affiliation{Center for Interdisciplinary Exploration and Research in Astrophysics,
Northwestern University, Evanston, IL 60208, USA}
\author{Shelby Klomp}
\affiliation{Center for Fundamental Physics, Department of Physics and Astronomy,
Northwestern University, Evanston, IL 60208, USA}
\author{Jackson Larsen}
\affiliation{Department of Physics and Astronomy, University of California, Davis, California 95616, USA}
\author{Jacob Sprague}
\affiliation{Center for Interdisciplinary Exploration and Research in Astrophysics,
Northwestern University, Evanston, IL 60208, USA}
\author{Zhiyuan Wang}
\affiliation{Center for Fundamental Physics, Department of Physics and Astronomy,
Northwestern University, Evanston, IL 60208, USA}
\author{George Winstone}
\affiliation{Center for Fundamental Physics, Department of Physics and Astronomy,
Northwestern University, Evanston, IL 60208, USA}
\author{Maddox Wroblewski}
\affiliation{Center for Fundamental Physics, Department of Physics and Astronomy,
Northwestern University, Evanston, IL 60208, USA}
\affiliation{Current Affiliation: Department of Physics, University of Texas at Austin,
Austin, TX 78712, USA}

\author{Shane L.\ Larson}
\affiliation{Department of Physics, Clarkson University, Potsdam, NY 13699, USA}
\author{Andrew A. Geraci}
\affiliation{Center for Fundamental Physics, Department of Physics and Astronomy,
Northwestern University, Evanston, IL 60208, USA}
\affiliation{Center for Interdisciplinary Exploration and Research in Astrophysics,
Northwestern University, Evanston, IL 60208, USA}
\author{Nancy Aggarwal}
\email{nqaggarwal@ucdavis.edu}
\affiliation{Department of Physics and Astronomy, University of California, Davis, California 95616, USA}

\date{\today}

\begin{abstract}
    Optically levitated sensors inside a Fabry–Pérot cavity have been proposed for high-frequency gravitational-wave (GW) detection, though their configuration for gravitational wave sensitivity exhibits counterintuitive features. We provide a new detailed general relativistic derivation of the interaction between a gravitational wave and a levitated object in an optical cavity, demonstrating gauge independence of the observable response. We find a strong asymmetric dependence of the strain signal on trap position, maximized when the sensor is located near the input mirror, and provide an in-depth explanation of its origin from multiple gauge perspectives. A key new result of this work is the consequence of this asymmetry on the noise coupling: the coupling of input-mirror displacements to the strain signal can be highly suppressed relative to that of end-mirror displacements and common-mode mirror motion. These results clarify the physical origin of the gravitational wave interaction with such a sensor and establish crucial design principles for optical levitation based high-frequency GW detectors.

\end{abstract}

\maketitle

\section{Introduction}

The direct detection of gravitational waves (GWs) by ground-based laser interferometers \cite{LIGOScientific:2018mvr, LIGOScientific:2020ibl, KAGRA:2021vkt, LIGOScientific:2025hdt} and the recent evidence for a stochastic GW background from pulsar timing arrays (PTAs) \cite{NANOGrav:2023gor, EPTA:2023gyr} have opened a new observational window onto the universe. Numerous detector concepts are currently being developed to extend GW sensitivity across a broad frequency landscape, targeting complementary bands and source populations \cite{ET:2025xjr, Hall2022,  LISA:2017pwj, Kawamura:2020pcg, MAGIS-100:2021etm, Takano:2024vht}. The ultra-high-frequency (UHF) band, however, spanning roughly tens of kilohertz to megahertz and beyond, remains largely unprobed, yet harbors a rich landscape of potential sources including scalar and vector boson clouds, cosmological first-order phase transitions, oscillating string networks, primordial black hole mergers, and exotic compact binary systems \cite{aggarwal2021challenges,aggarwal2025challenges,SpragueAxionClouds}. Developing detector concepts capable of reaching this band is therefore a pressing goal for the field.

Among the most promising proposals for UHFGW detection are compact detectors employing optically levitated sensors inside Fabry–Pérot cavities, which exhibit tunable detection frequency in the UHFGW band \cite{LSD2012,LSD2020}, especially well-suited to searching for scalar and vector boson clouds \cite{SpragueAxionClouds}. Such detectors represent a natural application of cavity optomechanics to GW detection, leveraging the tunability and ultralow mechanical dissipation of levitated oscillators to access the UHFGW band. In such a scheme, a dielectric particle or mirror is trapped at the focus of an intracavity optical field, and the displacement of the trapped object relative to the cavity is read out interferometrically. A network of these detectors is currently under construction in Evanston, Illinois and  Davis, California, making a rigorous theoretical foundation for their operation timely. Prior work \cite{LSD2012} has shown that such a system exhibits a nontrivial and asymmetric dependence of the GW strain signal on the position of the trap within the cavity — a result which is counterintuitive from the perspective of standard interferometer physics and has not yet been carefully explained in the literature.

In this paper, we examine this asymmetry in detail, carrying out the first self-contained and completely relativistic derivation of the relative displacement induced by a passing GW between a levitated object and its trap site.
We clarify the physical origins of the asymmetry and demonstrate that it has important considerations for how the GW signal and detector noise will impact the sensitivity of the instrument. In particular, we identify a strong suppression of input-mirror displacement noise compared to end-mirror displacement and common-mode length noise at sufficiently low frequencies, with direct implications for detector design. Additionally, we extend the fully relativistic calculation to general frequencies and highlight how the hierarchy in the signal couplings to GWs and cavity length noises can evolve drastically at frequencies above the cavity linewidth.

The interaction of a GW with a composite system — one in which an optically trapped object is coupled to two cavity mirrors, each of which also responds to the GW — is subtle. The result depends on how one treats the motion of masses and the behavior of the electromagnetic field in a curved spacetime background, and intermediate steps in such calculations are manifestly gauge dependent. Moreover, gauge choice and frame choice, such as which mirror is placed at the coordinate origin, are distinct, though they are often conflated in the literature. Demonstrating that the final observable is gauge independent and coordinate invariant is therefore a non-trivial and necessary part of establishing the result on formal footing. 

The present paper provides this derivation in full, working explicitly in two gauges — the transverse-traceless (TT) gauge and the local Lorentz (LL) gauge — and demonstrates that the physically observable signal is identical in both and independent of the choice of coordinate origin and cavity orientation. Because of the non-intuitive form of the trap site asymmetry at face value and its crucial bearing on optimal detector configurations, an in-depth discussion on its source with fully relativistic considerations is warranted. Therefore, in addition to the derivation itself, we include detailed explanations for the generation of this asymmetry from the perspectives of both the TT and LL gauges. In both gauges, it fundamentally originates from a fixed phase relation between the counter-propagating waves in the cavity which is imposed exclusively by the end mirror.\footnote{In this paper, we define the ``input mirror" as the mirror through which the laser enters a Fabry-Pérot cavity, and the ``end mirror" as the other mirror in the cavity, furthest from the laser source.}

A related subtlety at high GW frequencies is the long-wavelength approximation, in which the GW wavelength is assumed much larger than the detector. While this approximation holds comfortably  for well-studied detectors such as LIGO, Virgo, and PTAs, it becomes less reliable as GW frequencies increase into the UHF regime, where the wavelength can become comparable to detector dimensions. To avoid confirming gauge invariance only trivially, we will analytically compute the relative displacement to first order in the dimensionless quantity $\Omega L/c$, where $\Omega$
is the angular frequency of the GW and $L$ is the detector length, one order beyond the standard long-wavelength approximation.
Then, towards the end of the paper, we will numerically compute results valid to all orders in $\Omega L/c$.

This foundational contribution is timely beyond the context of levitated sensors alone. Full relativistic analyses of GW detector designs have been carried out in recent years for linear bar detectors \cite{weber1960detection}, large interferometers \cite{gerstenshtein1963detection,braginskii1965gravitational,cooperstock1993laser,
Schilling1997,rakhmanov2004characterization,RakhmanovLLGauge,tarabrin2007interaction,melissinos2010response, Koop:2013vxa, Essick:2017wyl, Rakhmanov:2008is, Li:2022mvy, Jungkind:2025oqm}, PTAs \cite{Deng:2010ut, McGrath:2021tia, Boitier:2020xfx}, atom interferometers \cite{Schaffrath:2025rsn, Wang:2021hrg}, microwave cavities \cite{Berlin:2021txa}, and torsion bar detectors \cite{Nakamura:2014wia}, but optically levitated sensors are a notable example of a system that has not yet received such treatment. As the GW detector landscape expands rapidly, with a growing number of proposed concepts operating at high frequencies, careful treatment of gauge and reference frame dependence is becoming increasingly important. Many of these systems cannot be idealized as freely falling test masses in the TT gauge — the standard simplification that underlies most GW detector analyses — and applying that framework uncritically can lead to errors. We hope the explicit two-gauge treatment presented here serves as a methodological reference for the broader community working on UHFGW detector design and analysis.

The remainder of this paper is organized as follows. In Sec.~\ref{sec: preliminaries}, we briefly outline the formalism used to describe the electric field propagating in the experimental model. Working to first order $\Omega L/c$, we prove in Secs.~\ref{sec: TT} and \ref{sec: LL} that both the TT and LL gauges predict an invariant relative displacement between the cavity antinode and the levitated particle, demonstrating gauge independence and coordinate invariance. This gauge-invariant displacement features an interesting asymmetry dictated by the position of the trap relative to the input mirror — a result reported previously but explained here for the first time — and in Sec.~\ref{sec: asymmetry origin} we offer a physical interpretation of its origin. In Sec.~\ref{sec: beyond first order}, we extend our results to general values of $\Omega L/c$, identifying where established intuition breaks down at high frequencies. In Sec.~\ref{sec: noise implications}, we discuss the implications of these results --- most notably, the trap site asymmetry --- on the propagation of noise within a detector and resulting considerations for detector design. Finally, we conclude in Sec.~\ref{sec: conclusion}.

\begin{figure}
    \centering
    \includegraphics[width=0.95\linewidth]{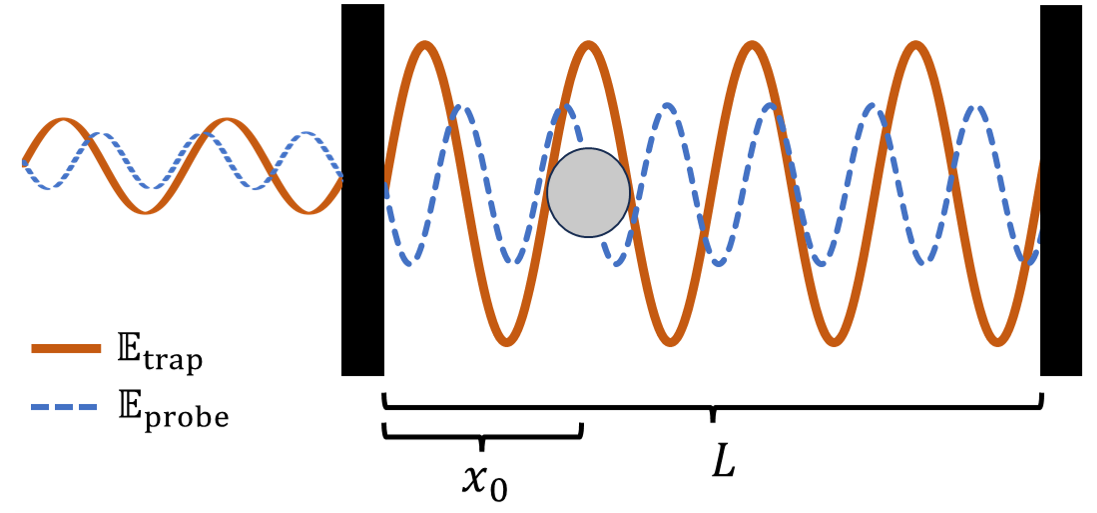}
    \caption{Schematic diagram of a single arm of a levitated-sensor-based GW detector. A trap beam (solid) with frequency $\omega_0$ and weak detuned probe beam (dashed) enter the cavity of length $L$ from the left. A dielectric nanoparticle is trapped a distance $x_0$ from the input mirror, at an antinode of the trap laser's standing wave in the cavity. The probe beam reads out the sensor position. Diagram not to scale.}
    \label{fig: experiment diagram}
\end{figure}

\section{The effect of a gravitational wave on a cavity optical trap}
\label{sec: full GR antinode displacement}

We now compute the effect of an incident GW on an optical trap inside a Fabry-Pérot cavity, choosing our coordinates so that the cavity is aligned with the $x-$direction and lies at $y=z=0$. 

In Fig. \ref{fig: experiment diagram}, we present a basic diagram of this class of detector. The experiment operates using two coaxial lasers: a strong trap beam with frequency $\omega_0$ and a weak detuned probe beam. The trap beam levitates a dielectric nanoparticle a distance $x_0$ from the cavity input mirror, in the manner of an optical tweezers \cite{Ashkin:1978zsu, Ashkin:86}. When the nanoparticle is sub-wavelength in size, it is trapped at an antinode of the trap beam standing wave in the cavity, where the potential due to the interaction between the electric field and the polarizable material is minimized \cite{ChangSpheres, ChangThinDiscs}. The probe beam primarily reads out the position of the sensor along the cavity axis.

The reader should note that we assume the sensor's dimensions are much smaller than the trap light wavelength $\lambda_0$; thus its insertion into the cavity does not noticeably disrupt the global standing wave configuration.

As a GW enters the system, the phase accumulated by light over free space evolves so that the minimum of the potential (i.e., the antinode of the standing wave) moves relative to the levitated sensor. Since the potential near the minimum is well-described by that of a harmonic oscillator, the resulting driving force is essentially linear in the displacement of the trap site (antinode) relative to the levitated sensor. Thus, this relative displacement is a useful proxy for contextualizing the full experimental response to a GW. For the sake of simplicity, we consider a plus-polarized plane GW with strain $h(t)\ll 1$ which propagates along the $z-$axis and whose polarization is aligned with the $x-$axis; general polarizations and propagation directions can be analyzed with slightly more complexity in the mathematics.

\subsection{Preliminaries}
\label{sec: preliminaries}
We assume an electric field $\vec{\mathbb{E}}$ which is polarized purely in the $z-$direction:
\begin{equation}
    \vec{\mathbb{E}}(x,y,z,t)=f(y,z)E(x,t)\hat z,
\end{equation}
where $f(y,z)$ is the cross-sectional distribution of the field amplitude. We will neglect the cross-sectional profile for the remainder of this derivation, as it is only the variation of the field along the cavity axis which provides the important physics here.

\begin{figure}
    \centering
    \includegraphics[width=8cm]{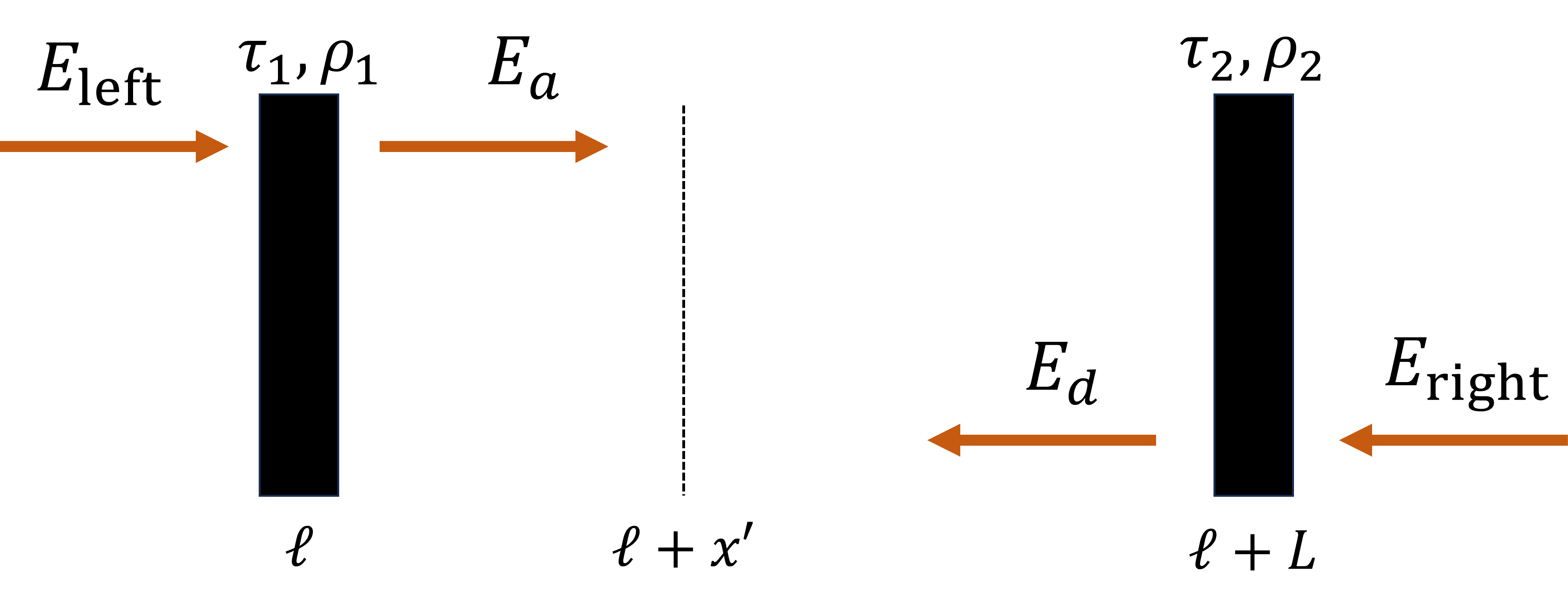}
    \caption{The fields inside and outside of a simple Fabry-Pérot optical cavity. When the laser enters from the left, $E_\text{left}\not=0$ and $E_\text{right}=0$, and vice-versa when entering from the right.}
    \label{fig:fields diagram}
\end{figure}

Our cavity toy model is pictured in Fig. \ref{fig:fields diagram}. Two mirrors of amplitude reflectivity and transmissivity $\rho_1$, $\rho_2$, $\tau_1$, and $\tau_2$ are located at coordinates $\ell$ and $\ell+L$, and the dielectric particle is excluded from the toy model due to its small size. The translation by $\ell$ allows us to test whether our results are invariant under the choice of coordinate origin. The right and left-propagating components of the fields at the interior surfaces of the infinitesimally thin cavity mirrors are labeled as $E_a$ and $E_d$, respectively. Meanwhile, $E_{\text{left,right}}$ are the input electric fields incident from the left and right of the cavity; they evolve as $\sim\exp(-i\omega_0t)$, where $\omega_0=2\pi c/\lambda_0$ is the light carrier frequency. We aim to compute the electric field at any coordinate $\ell+x'$ within the cavity.

Over a length of free space, the electric field at the far end is related to that of the near end by the time delay of the light propagating over that space. In the case where the field only enters from the left, this is manifest as 
\begin{subequations}
    \begin{equation}
        E_a(t)=\tau_1E_\text{left}(t)-\rho_1E_d(t-T(\ell+L,\ell)[t]),
    \end{equation}
    \begin{equation}
        E_d(t)=-\rho_2E_a(t-T(\ell,\ell+L)[t]).
    \end{equation}
    \label{eq: full time delayed rotation}
\end{subequations}
Here, $T(\ell+L,\ell)[t]$ and $T(\ell,\ell+L)[t]$ are defined as the time for light to propagate from the end mirror to the input mirror and from input mirror to end mirror, respectively, arriving at time $t$ and accounting for all gravitational effects. When the cavity is unperturbed, both quantities are equal to $L/c$.

This recursive set of equations can be solved in terms of an infinite series:
\begin{equation}
    E_a(t)=\tau_1\sum_{n=0}^\infty (\rho_1\rho_2)^nE_\text{left}(t-\sum_{j=0}^nT_j),
\end{equation}
where $T_j$ is the $j$-th previous round trip travel time, i.e.
\begin{equation}
    \begin{aligned}
        &T_0=0,
        \\
        &T_1=T(\ell+L,\ell)[t]+T(\ell,\ell+L)[t-T(\ell+L,\ell)[t]],
        \\
        &T_2=T(\ell+L,\ell)[t-T_1]
        \\
        &\quad\quad+T(\ell,\ell+L)[t-T_1-T(\ell+L,\ell)[t-T_1]],
    \end{aligned}
\end{equation}
and so on. Fortunately, when solving for fields to linear order in the GW strain, it is sufficient to insert unperturbed values into the bracketed arguments of the $T$ function, as $\mathcal{O}(h)$ corrections to those arguments result in $\mathcal{O}(h^2)$ corrections to the total roundtrip times. Therefore, we can safely approximate
\begin{equation}
    \begin{aligned}
        &T_1\approx T(\ell+L,\ell)[t]+T(\ell,\ell+L)[t-L/c],
        \\
        &T_2\approx T(\ell+L,\ell)[t-2L/c]+T(\ell,\ell+L)[t-3L/c],
        \\
        &\ldots
        \label{eq: T_m}
    \end{aligned}
\end{equation}

To solve for the internal fields in full generality, we must know the entire history of the GW strain, as $E_a(t)$ depends roughly on $E_a(t-2L/c)$, which itself depends roughly on $E_a(t-4L/c)$, and so on. Taking incident perturbations to oscillate at frequency $\Omega$, we will address this problem by assuming that the perturbing signal has persisted stably for much longer than $1/\kappa$, where $\kappa$ is the cavity decay rate. This time scale is on the order of the cavity round-trip time multiplied by its finesse and ensures that the field evaluation retarded times $t-L/c$, $t-2L/c$, etc. all reference a coherent perturbing signal, at least up to the point where the compounded reflection coefficient $\rho_{1,2}<1$ makes further contributions from previous reflected fields safely negligible.

After obtaining $E_a(t)$ and $E_d(t)$, we can finally solve for the field at all points inside the cavity:
\begin{multline}
    E(x',t)=E_a(t-T(\ell,\ell+x')[t])
    \\
    +E_d(t-T(\ell+L,\ell+x')[t]),
\end{multline}
where $T(\ell,\ell+x')[t]$ and $T(\ell+L,\ell+x')[t]$ are the light propagation times starting at the input or end mirror, respectively, and arriving at \textit{coordinate} $x'$ at time $t$.

We locate the antinodes by finding the maxima of the squared electric field amplitude $EE^*$, which given the form of $E_\text{left}(t)$ effectively averages over the carrier light period. This implicitly assumes that the GW frequency $\Omega$ is much less than $\omega_0$, though this assumption is quite reasonable even in the UHFGW regime. Locating the antinodes for the unperturbed (no GW) and perturbed system allows for the computation of $\delta x_\text{antinode}$. After computing $\delta x_\text{sensor}$, the GW-induced displacement for a sensor whose unperturbed position is $\ell+x_0$, we can extract $\delta x_\text{antinode-sensor}=\delta x_\text{a-s}\equiv\delta x_\text{antinode}-\delta x_\text{sensor}$.

We wish to show that $\delta x_\text{antinode-sensor}$ is invariant by computing it in both the transverse-traceless (TT) and local Lorentz (LL) gauges and two different reference frames: one where the laser enters from the left side of the cavity, and the other where the laser enters from the right side. The form of the electric field where $E_\text{right}\not=0$, $E_\text{left}=0$ can be inferred straightforwardly from the previous expressions in this section. The calculation of $T(a,b)[t]$ is different between the TT and LL gauges, and this is where the computations in the two gauges diverge from one another. 
We now proceed to the details of these calculations.

\subsection{Transverse-traceless gauge}
\label{sec: TT}
In the TT gauge, all free masses remain at the same comoving coordinates as a GW passes through the system\footnote{We assume the GW frequency exceeds all mechanical resonance frequencies in the cavity mirror mounts so that the cavity mirrors behave like free masses.}. The distance between objects at fixed comoving coordinates, however, changes through perturbations to the spacetime metric. For our designated GW configuration, the metric in Cartesian coordinates is $g_{\mu\nu}=\text{diag}\{-1,1+h(t),1-h(t),1\}$. 

Since photons propagate along null geodesics, their trajectory parametrized by affine parameter $\xi$ must satisfy
\begin{equation}
    -c^2\frac{d^2t}{d\xi^2}+(1+h(t))\frac{d^2x}{d\xi^2}=0.
\end{equation}
Fulfilling this condition amounts to a description of the evolution of the coordinate time as light propagates through free space:
\begin{equation}
    \frac{dt}{dx}=\frac 1c\sqrt{1+h(t)}\approx\frac1c(1+\frac12h(t)),
\end{equation}
with the approximation carried out to first order in $h$.

Therefore, in the TT gauge, we find 
\begin{equation}
\begin{gathered}
    T^\text{TT}(a,b)[t]=\frac{1}{c}~\Bigr|\int_{a}^{b}\Bigr(1+\frac{h(t_r)}{2}\Bigr)dX\Bigr|,
\end{gathered}
\end{equation}
where $t_r=t-|b-X|/c$ is the light propagation retarded time\footnote{We do not need to correct for the perturbed velocity of light in $t_r$ because it results in overall $\mathcal{O}(h^2)$ corrections to the phase. \label{note: coordinate velocity}}. Here, we assume time dependence in the strain of the form \begin{equation}
    h(t)=h_0\cos(\Omega t),
\end{equation} 
which leads to the following time segments:
\begin{subequations}
\begin{align}
\begin{split}
    &T^\text{TT}(\ell,\ell+L)[t]=T^\text{TT}(\ell+L,\ell)[t]=\\&\quad
    \frac{L}c+\frac{h_0}{2\Omega}\Bigr(\sin(\Omega t)
    -\sin(\Omega t-\frac{\Omega L}{c})\Bigr),
\end{split}
\\
\begin{split}
    &T^\text{TT}(\ell,\ell+x')[t]=\\&\quad\frac{x'}c+\frac{h_0}{2\Omega}\Bigr(\sin(\Omega t)
    -\sin(\Omega t-\frac{\Omega x'}{c})\Bigr),
\end{split}
\\
\begin{split}
    &T^\text{TT}(\ell+L,\ell+x')[t]=\\&\quad\frac{L-x'}c+\frac{h_0}{2\Omega}\Bigr(\sin(\Omega t)
    -\sin(\Omega t-\frac{\Omega (L-x')}{c})\Bigr).
\end{split}
\end{align}
\end{subequations}
With these times calculated, we can extract the delayed round trip times $T_m$ and finally compute $E(\ell+x',t)$. Prior to GW incidence and with the laser entering from the left, the antinodes appear where
\begin{equation}
    \frac{2(L-x_0)\omega_0}{c}=(2n+1)\pi,~n\in\mathbb{Z}\quad\text{($E_\text{right}=0$)}.
    \label{eq: unperturbed antinodes left}
\end{equation}
There is a subtle point hidden in this expression, namely that changes in the cavity length $L$ due to DC displacements of the input and end mirror are not equivalent. If the cavity length increases due to a displacement of the input mirror to the left by $\delta L$, $x_0$ must increase by $\delta L$. However, $x_0$ is measured from the surface of the input mirror (at DC), so in the laboratory frame, the end mirror and antinode locations do not move. However, if the same length increase were facilitated by a displacement of the end mirror to the right, $x_0$ must increase again, but now both the end mirror and antinode locations move in the laboratory frame. 

This asymmetry persists when considering the setup where the laser enters from the right. In this case, the unperturbed antinode locations satisfy
\begin{equation}
    \frac{2x_0\omega_0}{c}=(2n+1)\pi,~n\in\mathbb{Z}\quad\text{($E_\text{left}=0$)}.
    \label{eq: unperturbed antinodes right}
\end{equation}
The asymmetry has important implications on how certain noise sources propagate to readout channels, which we will discuss further in Sec. \ref{sec: noise implications}. The origins of the asymmetry itself will be addressed in Sec. \ref{sec: asymmetry origin}.

After expanding the field solutions to linear order in $h$ as well as $\Omega L/c$ and $\Omega x'/c$, we extract the displacement of the field maxima, $\delta x_\text{antinode}$. Finally, since the levitated sensor is a free mass (for the sake of computing the force from the trap laser), $\delta x_\text{sensor}=0$. Within the TT gauge, we consider the two reference frames where the trap laser enters from the left and right side of the cavity. When it enters from the left ($E_\text{right}=0$), we find
\begin{equation}
    \delta x^\text{TT}_\text{a-s}=\frac{h_0(L-x_0)}{2}\Bigr(\cos\Omega t-\frac{\Omega L}{c}\frac{|r|^2-1}{|r-1|^2}\sin\Omega t\Bigr),
    \label{eq: delta x antinode sensor left}
\end{equation}
where $r\equiv\rho_1\rho_2\exp(2i\omega_0L/c)$,
whereas when the trap laser enters from the right,
\begin{equation}
    \delta x^\text{TT}_\text{a-s}=-\frac{h_0x_0}{2}\Bigr(\cos\Omega t
    \\
    -\frac{\Omega L}{c}\frac{|r|^2-1}{|r-1|^2}\sin\Omega t\Bigr).
    \label{eq: delta x antinode sensor right}
\end{equation}
Note that the mirror coefficients $\rho_1$ and $\rho_2$ only enter at first order, when the phase offset of the GW-produced sideband fields becomes relevant. 

In both cases, the magnitude of the displacement displays an asymmetric dependence on the sensor location, scaling with the distance of the sensor from the end mirror. This agrees with the zeroth order calculation in \cite{LSD2012} that the separation between the sensor and cavity antinode induced by the GW is maximized when the sensor is levitated near the input mirror.

\subsection{Local Lorentz gauge}
\label{sec: LL}
In the local Lorentz gauge, coordinates remain fixed (i.e., they are not comoving), while the leading order effect of a GW is to move free masses in space -- a free mass at coordinate $X$ is displaced by $\frac12h(t)X$.

In fact, the accumulated field phase experiences $\mathcal{O}(h)$ perturbations due to three sources: not only the displacement of the free mirrors, but also the modified rate of ticking clocks at a fixed observing coordinate and the modified coordinate velocity of light. It is shown in \cite{RakhmanovLLGauge} that all three effects need to be accounted for to retain gauge invariance in the phase of light accumulated during the round trip of a photon through a cavity when a GW is incident. Note that \cite{RakhmanovLLGauge} describes the final two sources as ``localized" and ``distributed" gravitational redshift, respectively. Here, we aim to show that all three contributions are necessary to show gauge invariance in quantities which depend on the field structure within the cavity, not just for the phase accumulated in a full round trip.

We write the total accumulated time between points as $\delta T^\text{LL}\equiv T^\text{LL}_{h=0}+\delta T^\text{LL}_x+\delta T^\text{LL}_t+\delta T^\text{LL}_v$ where $T^\text{LL}_{h=0}$ is the time accumulated without a GW present and the remaining terms are the three sources of perturbations.

We first consider the effect of the mirror displacement $\delta T_x$. Using the same conventions as in the TT gauge calculation, the propagation time perturbations from the mirror displacement are
\begin{subequations}
\begin{align}
\begin{split}
    &\delta T^\text{LL}_x(\ell,\ell+L)[t]=\frac{1}{c}\Bigr[\frac12(\ell+L)h(t)-\frac12\ell h(t-L/c)\Bigr],
\end{split}
\\
\begin{split}
    &\delta T^\text{LL}_x(\ell+L,\ell)[t]=\frac{1}{c}\Bigr[\frac12(\ell+L)h(t-L/c)-\frac12\ell h(t)\Bigr],
\end{split}
\\
\begin{split}
    &\delta T^\text{LL}_x(\ell,\ell+x')[t]=\frac{1}{c}\Bigr[-\frac12\ell h(t-x'/c)\Bigr],
\end{split}
\\
\begin{split}
    &\delta T^\text{LL}_x(\ell+L,\ell+x')[t]=\frac{1}{c}\Bigr[\frac12(\ell+L)h(t-(L-x')/c)\Bigr].
\end{split}
\end{align}
\end{subequations}

Next, the ``localized redshift" emerges from the fact that measurements need to be carried out in the proper time of a stationary observer, whereas the previous terms are computed in coordinate time. The rate of ticking clocks at coordinate $x$, $dt^*(x)$, relative to coordinate time $dt$ is given by 
\begin{equation}
    dt^*(x)/dt=\sqrt{-g_{00}(x,t)}.
\end{equation}
In the LL gauge, the metric is given by \cite{RakhmanovLLGauge}
\begin{equation}
    g_{\mu\nu}^\text{LL}=\eta_{\mu\nu}-\frac{2}{c^2}\Phi
    \begin{pmatrix}
        1 & 0 & 0 & 1
        \\
        0 & 0 & 0 & 0
        \\
        0 & 0 & 0 & 0
        \\
        1 & 0 & 0 & 1
    \end{pmatrix},
\end{equation}
where (in our chosen coordinate system) $\Phi\equiv-\frac14\ddot h(t)x^2$. The localized redshift this gives an $\mathcal{O}(h)$ correction of
\begin{subequations}
\begin{align}
\begin{split}
    &\delta T^\text{LL}_t(\ell,\ell+L)[t]=\frac{1}{c^2}\int_{t-L/c}^{t}\Phi(\ell+L,t')dt',
\end{split}
\\
\begin{split}
    &\delta T^\text{LL}_t(\ell+L,\ell)[t]=\frac{1}{c^2}\int_{t-L/c}^{t}\Phi(\ell,t)dt',
\end{split}
\\
\begin{split}
    &\delta T^\text{LL}_t(\ell,\ell+x')[t]=\frac{1}{c^2}\int_{t-x'/c}^{t}\Phi(\ell+x',t')dt',
\end{split}
\\
\begin{split}
    &\delta T^\text{LL}_t(\ell+L,\ell+x')[t]=\frac{1}{c^2}\int_{t-(L-x')/c}^{t}\Phi(\ell+x',t')dt'.
\end{split}
\end{align}
\end{subequations}

Finally, the modification to the coordinate velocity of light can be computed in the same manner as was done in the TT gauge calculation. We find
\begin{subequations}
\begin{align}
\begin{split}
    &\delta T^\text{LL}_v(\ell,\ell+L)[t]
    \\&\quad
    =-\frac{1}{c^3}\int_{\ell}^{\ell+L}\Phi\Bigr(X,t-\frac{L+\ell-X}{c}\Bigr)dX,
\end{split}
\\
\begin{split}
    &\delta T^\text{LL}_v(\ell+L,\ell)[t]
    \\&\quad
    =-\frac{1}{c^3}\int_{\ell}^{\ell+L}\Phi\Bigr(X,t-\frac{X-\ell}{c}\Bigr)dX,
\end{split}
\\
\begin{split}
    &\delta T^\text{LL}_v(\ell,\ell+x')[t]
    \\&\quad
    =-\frac{1}{c^3}\int_{\ell}^{\ell+x'}\Phi\Bigr(X,t-\frac{x+\ell-X}{c}\Bigr)dX,
\end{split}
\\
\begin{split}
    &\delta T^\text{LL}_v(\ell+L,\ell+x')[t]
    \\&\quad
    =-\frac{1}{c^3}\int_{\ell+x'}^{\ell+L}\Phi\Bigr(X,t-\frac{X-x-\ell}{c}\Bigr)dX.
\end{split}
\end{align}
\end{subequations}
Note that we can integrate to the unperturbed coordinates of the cavity mirrors, as corrections to the integration bounds will produce $\mathcal{O}(h^2)$ corrections overall.

Once again considering an incident plus-polarized GW with $h(t)=h_0\cos(\Omega t)$, we find (to first order in $\Omega L/c$)
\begin{subequations}
\begin{align}
\begin{split}
    &T^\text{LL}(\ell,\ell+L)[t]
    \\&\quad
    =\frac{L}{c}\Bigr(1+\frac {h_0}{2}\Bigr(\cos\Omega t-\frac{\Omega\ell}{c}\sin\Omega t\Bigr)\Bigr),
\end{split}
\\
\begin{split}
    & T^\text{LL}(\ell+L,\ell)[t]
    \\&\quad
    =\frac{L}{c}\Bigr(1+\frac {h_0}{2}\Bigr(\cos\Omega t+\frac{\Omega(\ell+L)}{c}\sin\Omega t\Bigr)\Bigr),
\end{split}
\\
\begin{split}
    & T^\text{LL}(\ell,\ell+x')[t]
    \\&\quad
    =\frac{x'}{c}\Bigr(1-\frac{h_0}{2}\frac{\Omega\ell}{c}\sin\Omega t\Bigr)-\frac{\ell}{c}\frac {h_0}{2}\cos\Omega t,
\end{split}
\\
\begin{split}
    &T^\text{LL}(\ell+L,\ell+x')[t]
    \\&\quad
    =\frac{L-x'}{c}\Bigr(1+\frac{\Omega(L+\ell)}{c}\frac {h_0}{2}\sin\Omega t\Bigr)
    \\
    &\hspace{2.92cm}+\frac{L+\ell}{c}\frac {h_0}{2}\cos\Omega t. 
\end{split}
\end{align}
\end{subequations}

With these times obtained, we can compute the field at all points within the cavity and once again solve for the perturbed antinode locations
\begin{multline}
    \delta x^\text{LL}_\text{antinode}\approx\frac{h_0(L-x_0)}{2}\Bigr(\cos\Omega t-\frac{\Omega L}{c}\frac{|r|^2-1}{|r-1|^2}\sin\Omega t\Bigr)
    \\+\frac12h_0(\ell+x_0)\cos\Omega t,
    \label{eq: antinode disp LL gauge}
\end{multline} 
where the laser enters from the left, and
\begin{multline}
    \delta x^\text{LL}_\text{antinode}\approx-\frac{h_0x_0}{2}\Bigr(\cos\Omega t-\frac{\Omega L}{c}\frac{|r|^2-1}{|r-1|^2}\sin\Omega t\Bigr)
    \\+\frac12h_0(\ell+x_0)\cos\Omega t,
    \label{eq: antinode disp LL gauge right}
\end{multline}
where it enters from the right.
Unlike in the TT gauge, however, the sensor is displaced due to the GW --  that is, $\delta x^\text{LL}_\text{sensor}=\frac12h(t)(\ell+x_0)$.  Combining these results to obtain $\delta x^\text{LL}_\text{antinode-sensor}$, we find identical results to Eqs. \ref{eq: delta x antinode sensor left} and \ref{eq: delta x antinode sensor right} when the laser enters from the left and right, respectively. We thus confirm that regardless of the choice of gauge or origin of the  coordinate system, the GW-induced displacement of the sensor relative to the laser antinode is consistent.

\subsection{Origin of the asymmetry}
\label{sec: asymmetry origin}
The asymmetry in the sensor-antinode displacement (i.e., the monotonic increase in the net displacement as the sensor approaches the input mirror) deserves further intuitive explanation. In this section, we will provide such an explanation for both the TT and LL gauges.

The asymmetry ultimately revolves around the fact that the end mirror imposes a fixed phase relation between the forward and backward propagating waves within the cavity, while the input mirror does not. This can be seen through the cavity field equations: taking $\phi_{x'\rightarrow L\rightarrow x'}$ to be the accumulated phase from $\ell+x'$ to the end mirror coordinate and back to $\ell+x'$, and assuming $\phi_{x'\rightarrow L\rightarrow x'}$ varies slowly relative to $L/c$, the forward and backward propagating waves $E_f$ and $E_b$ at $\ell+x'$ satisfy
\begin{equation}
    E_b(\ell+x',t)=-\rho_2\exp(i\phi_{x'\rightarrow L\rightarrow x'})E_f(\ell+x',t).
\end{equation} 
At zeroth-order in $\Omega L/c$, $\phi_{x'\rightarrow L\rightarrow x'}=2\omega_0(L-x')/c\times (1+h(t)/2)$ in both the TT and LL gauges.

The phase of the resulting standing wave depends on the phase difference between the forward and backward propagating waves at a given point (i.e., antinodes where the forward and backward waves are in-phase, and nodes where they are out-of-phase). Under the assumption of slow variation of $\phi_{x'\rightarrow L\rightarrow x'}$, the end mirror marks where exactly half of $\phi_{x'\rightarrow L\rightarrow x'}$ has been accumulated. Thus, as $\phi_{x'\rightarrow L\rightarrow x'}$ evolves (slowly) in time, the antinode location responds to maintain $(N+1/2)\pi$ phase accumulation\footnote{The extra $\pi/2$ comes from the additional $\pi$ phase imposed at the reflecting surface.} to and from the end mirror --- the standing wave is anchored to the end mirror, in a sense.

Notably, the input mirror \textit{does not} impose an analogous fixed phase relation between forward and backward propagating waves, as there is a contribution from the incoming light which interferes with the light already within the cavity that reflects off the input mirror. When the input mirror moves to the right, for example, the reflected light loses phase while the incoming light gains phase. The antinodes cannot be anchored to both mirrors, as this would lead to internal contradictions, so they must only be anchored to the end mirror.

\begin{figure}
    \centering
    \includegraphics[width=0.9\linewidth]{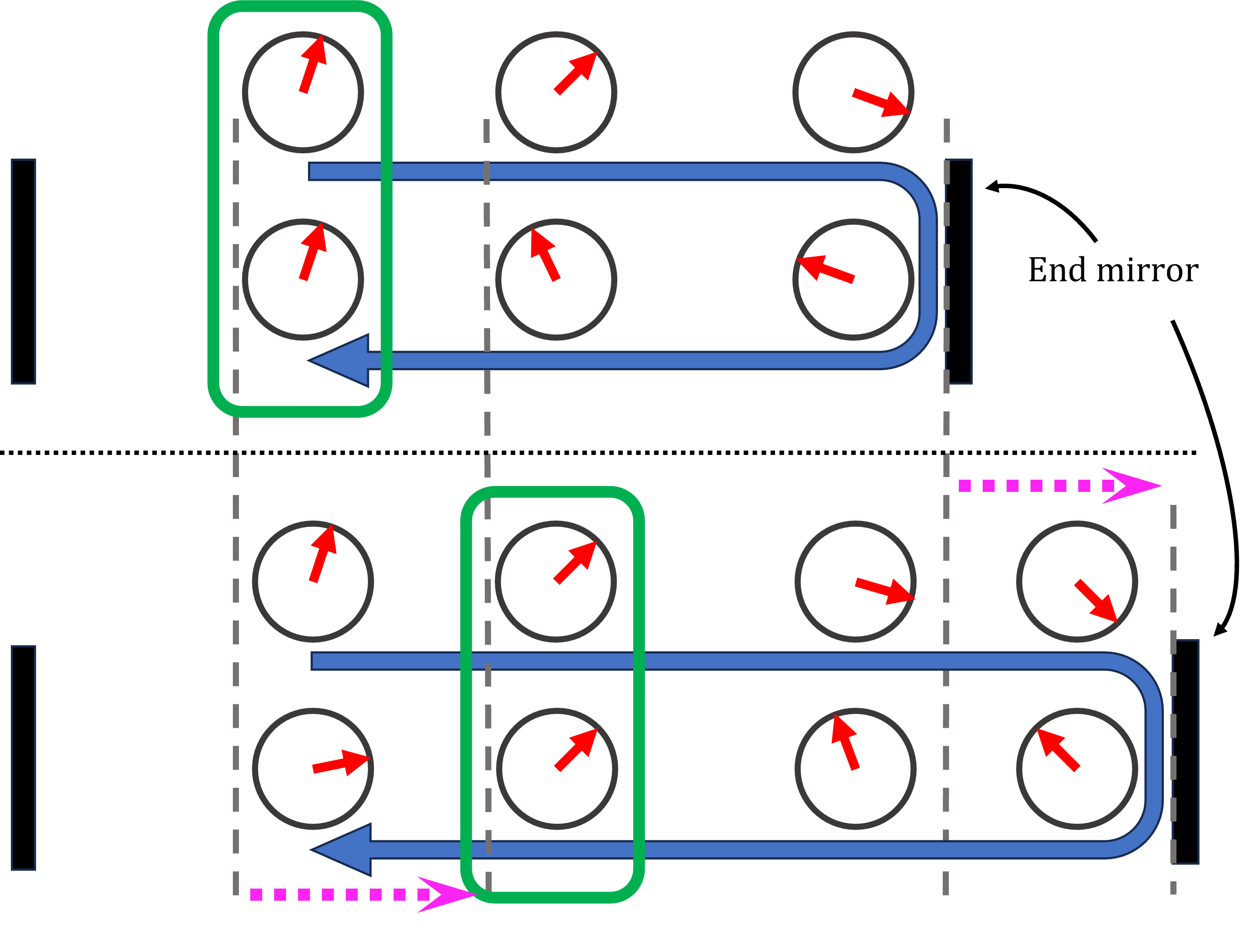}
    \caption{Phasors (red arrows) of the light field as evaluated along the blue space-like path to and from the end mirror, where spatial elements impose clockwise rotation and a phase shift of $\pi$ is imposed at the reflecting surface. Gray dashed lines indicate where the phasors are evaluated, and the pair of phasors which is in-phase (and thus denotes an antinode) is circled in green. In the bottom half of the image, the end mirror has been displaced by an amount given by the pink dashed arrow --- the antinode is displaced by an equal amount.}
    \label{fig:end mirror phasors}
\end{figure}

In Fig. \ref{fig:end mirror phasors}, we demonstrate this principle, applied to the LL gauge, using phasors. Recall that at in the LL gauge, the lowest order effect in $\Omega L/c$ is a displacement of the mirrors by $h(t)/2$ times their unperturbed separation from the coordinate origin --- changes to the coordinate velocity of light or the clock ticking rate only appear at higher orders. The light field is given a certain phase at the start of the blue path, and we track its evolution along this space-like path to and from the end mirror both before (top) and after (bottom) the end mirror is displaced. Before the end mirror is displaced, the phasors for the two counterpropagating fields align at the start of the blue path, indicating an antinode at that location. Once the end mirror is displaced, they no longer align at that point, but rather at a point shifted away from the input mirror by an amount identical to the end mirror displacement.

The anchoring of the standing wave to the end mirror location implies that in the LL gauge, each antinode should shift by $h(t)/2\times(L+\ell)$  when the laser enters from the left and by $h(t)/2\times \ell$ when the laser enters from the right. Indeed, both Eq. \eqref{eq: antinode disp LL gauge} and Eq. \eqref{eq: antinode disp LL gauge right} agree with this expectation at zeroth order in $\Omega L/c$. Importantly, these lowest order results are independent of $x_0$; thus, an incident GW displaces all antinodes by the same amount in this gauge, regardless of their unperturbed location along the cavity axis. When $h(t)>0$, the sensor moves to the left relative to the end mirror, and thus also relative to the antinode it is trapped at. The sensor displacement relative to the end mirror scales with the unperturbed distance between the objects, so the relative sensor-antinode displacement is maximized when the sensor is levitated as far from the end mirror as possible.

Meanwhile, in the TT gauge, the sensor's location is unmodified by the GW; instead, the coordinate velocity of light is reduced by a factor of $1+h(t)/2$. The frequency of the light is unmodified, as it is set by the laser source and $g_{tt}$ remains at 1 in this gauge. 
Consequently, the wavelength in coordinate space decreases, so the optical phase accumulates more rapidly with coordinate distance. In the phasor picture of Fig. \ref{fig:end mirror phasors}, this corresponds to the phasors rotating more rapidly in space.

The antinode locations correspond to where the forward and backward propagating waves differ in phase by $2\pi N$, $N\in\mathbb{Z}$. When the coordinate velocity is reduced (with $h(t)>0$) and the laser incident from the left, the increased spacial phase accumulation causes the antinodes to shift to the end mirror (to the right) to maintain the same phase difference. Furthermore, the displacement of each antinode must increase with its unperturbed distance from the end mirror, since the light traverses a longer path over  which the wave accumulates additional phase (i.e., the phasor accumulates additional rotation). Thus once again, the relative sensor-antinode displacement grows as the sensor is levitated farther from the end mirror.

It is worth noting that, assuming the cavity is locked on resonance prior to GW incidence, the strain shifts the cavity slightly off of its resonant configuration. This does decrease the circulating power in the cavity, but the decrease only appears at second order in the deviation of the round trip phase, and thus is negligible relative to the leading order effect of the antinode displacement.

\section{Beyond first order in $\Omega L/c$}
\label{sec: beyond first order}

Future iterations of these detectors will likely feature longer Fabry-Pérot cavities -- orders of magnitude longer than current prototype designs. For these configurations, the dimensionless $\Omega L/c$ can approach 1, necessitating a solution which does not treat this quantity as perturbative\footnote{Even when $L \sim c/\Omega$, at the GW frequencies probed here $\Omega\ll\omega_0$ is still valid.}. In this section, we compute the relative sensor-antinode displacement due to a GW or oscillatory mirror displacement for general values of $\Omega$. We do, however, retain the assumption that $h\ll 1$ and $\delta x_\text{m}/\lambda_0\ll 1$, where $\delta x_\text{m}$ is the mirror displacement magnitude, expanding the fields to first order in these quantities before numerically locating the displaced antinodes.

\subsubsection{Effect of a gravitational wave}
\label{sec: general freq gw}
\begin{figure}
    \centering
    \includegraphics[width=\linewidth]{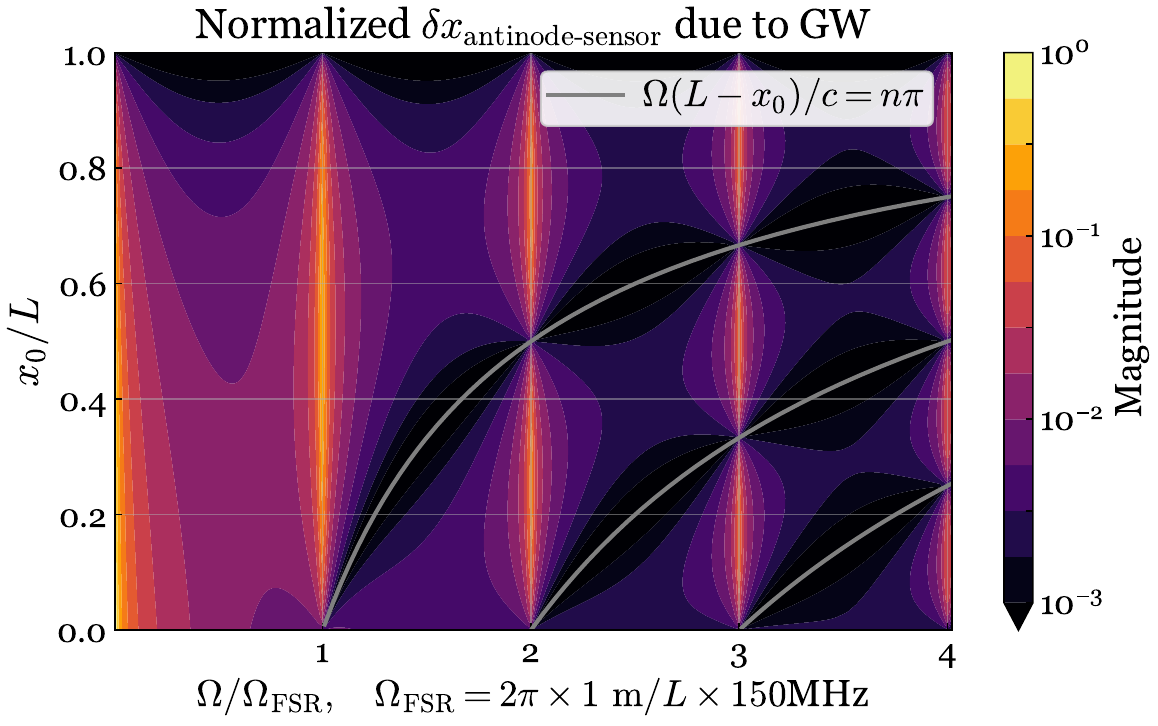}
    \caption{
    The magnitude of the relative sensor-antinode displacement due to a GW as a function of the rest position $x_0$ and GW frequency $\Omega$, normalized by $h_0L/2$, where we have set $\rho_2=\sqrt{0.96}$ and $r=0.96$. Large relative displacement occurs due to resonant sideband amplification at multiples of the free spectral range $\Omega_\text{FSR}$. Meanwhile, when $\Omega(L-x_0)/c=n\pi$, the displacement is zero.}
    \label{fig:gw displacement}
\end{figure}

In Fig. \ref{fig:gw displacement}, we plot the magnitude of the relative sensor-antinode displacement due to a GW as a function of its frequency and the sensor rest position\footnote{In practice, we adjust $x_0$ to the nearest antinode for each data point, but $L\gg \lambda_0$ so this effect is imperceptible in the plot.} $x_0$, normalized by the lowest-order cavity length change in the LL gauge, $h_0L/2$. The cavity is held on resonance with respect to $\omega_0$. Immediately evident are bands of large relative displacement at multiples of the cavity free spectral range (FSR). In these regions, the sideband fields generated by the GW at $\pm\Omega$ are also resonant in the cavity; the resulting $\mathcal{O}(h)$ contribution to the total electric field is amplified, enhancing the shift in the maximum of $EE^*$. 

We have already shown that at low frequencies, the relative displacement is maximized when the sensor is levitated as close to the input mirror as possible. When $\Omega$ reaches the scales of the FSR, however, this is not necessarily true. For example, for $\Omega=\Omega_\text{FSR}$, the maximum is at $x_0=L/2$. At this frequency, the carrier and sideband contributions to the total standing wave have the same phase at both mirrors. Their phase difference is therefore maximized in the middle of the cavity, so the displacement of the maximum of the total $EE^*$ is also maximized at this point for any given $h_0$.

Furthermore, the relative displacement falls to zero when $\Omega (L-x_0)/c=\pi N$, a result which cannot be obtained in the lowest-order approximation. The origin of this feature is the same of the zero in the response of a Fabry-Pérot Michelson interferometer to a GW at the FSR: when this condition is met, the light samples an entire oscillation of the GW propagating from the sensor to the end mirror and back, and thus the differential accumulated phase vanishes. In the phasor picture we described in Sec. \ref{sec: asymmetry origin}, then, the antinodes remain fixed. Therefore, much like the response of interferometric GW detectors at high frequencies, there exist frequencies at which a single arm with a levitated sensor is inherently insensitive to GWs incident from directions normal to that arm. These deviations from the low frequency behavior occur well above the targeted band of present-day proposals \cite{LSD2020}; nevertheless, they may prove important for future designs which target even greater frequencies.

\begin{figure}
    \centering
    \includegraphics[width=\linewidth]{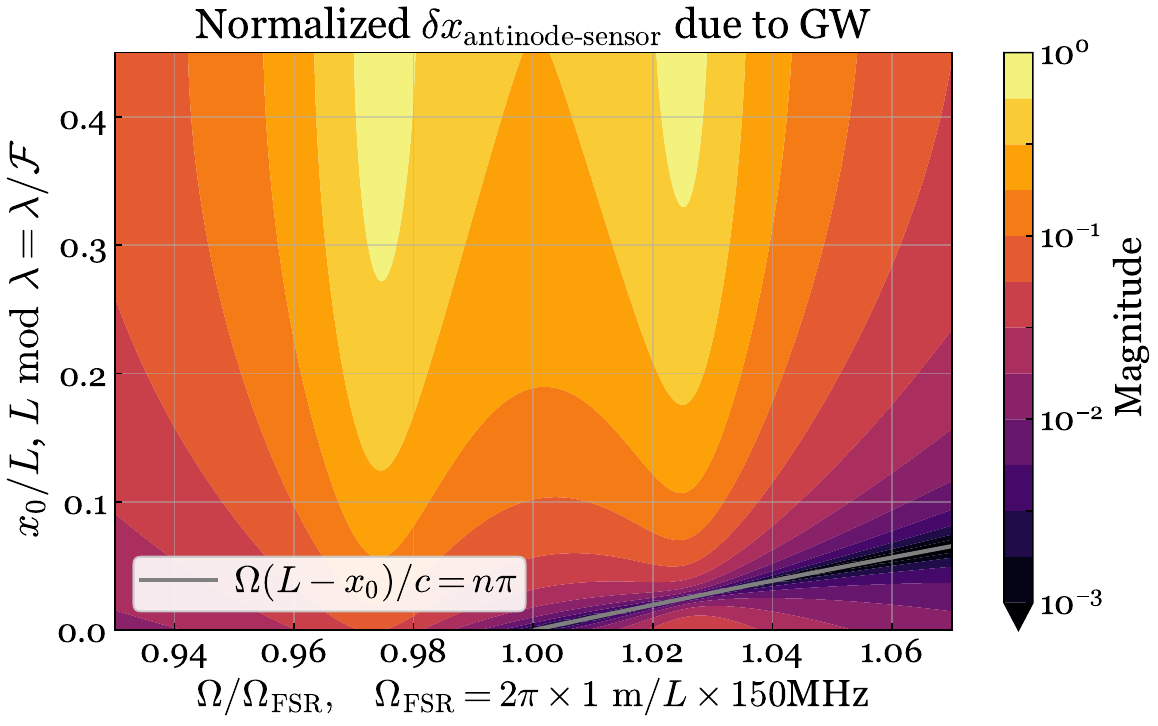}
    \caption{Identical to Fig. \ref{fig:gw displacement}, except we have detuned the cavity slightly so that $r=0.96\times\exp(4\pi i/\mathcal{F})$, where $\mathcal{F}\approx 77$ is the cavity finesse. We have also zoomed into a smaller region of $\Omega/\Omega_\text{FSR}$ and $x_0/L$ to highlight the two bands of amplified antinode displacement which appear where the GW-driven sidebands are resonant in the cavity.
    }
    \label{fig:gw displacement detuned}
\end{figure}

It is illustrative to also consider the GW-induced displacement when the cavity is held slightly off resonance with respect to the trap light. In Fig. \ref{fig:gw displacement detuned}, we compute the relative antinode displacement magnitude for the case where the cavity length is $1/\mathcal{F}$ times the trap wavelength off from resonance, where $\mathcal{F}$ is the cavity finesse.

We identify two frequency bands where the relative displacement is magnified. These bands are located symmetrically about integer multiples of the FSR and correspond to frequencies where the sideband fields produced by the GW at either $+\Omega$ or $-\Omega$ resonate in the cavity, amplifying the $\mathcal{O}(h)$ contribution to the total field.

With the cavity detuned, the sideband fields in these bands experience greater amplification in the cavity than they do in the low frequency limit. Therefore, it is possible for the antinode displacement magnitude to exceed $h_0L/2$, though from our numerical tests we find that this generally can only occur when the detuning well exceeds the cavity linewidth. Such large detunings are likely impractical for experimental operation, as the partial destructive interference of the carrier field in the cavity would necessitate amplified input power to achieve comparable optical trap frequencies.

\subsubsection{Effect of mirror motion}

End mirror motion in the laboratory frame $\delta x_\text{end}(t)=\delta x_\text{m}\cos(\Omega t)$ generates a $j$-th previous roundtrip time of
\begin{equation}
    T_j^\text{end}=\frac{2L}{c}+\frac{2\delta x_\text{m}}{c}\cos\Bigr(\Omega(t-\frac{(2j-1)L}{c})\Bigr)~~(j>0),
\end{equation}
whereas identical motion of the input mirror leads to
\begin{multline}
    T_j^\text{input}=\frac{2L}{c}-\frac{\delta x_\text{m}}{c}\Bigr[\cos\Bigr(\Omega(t-\frac{(2j-2)L}{c})\Bigr)\\
    +\cos\Bigr(\Omega(t-\frac{2jL}{c})\Bigr)\Bigr]~~(j>0).
\end{multline}

\begin{figure}
    \centering
    \includegraphics[width=\linewidth]{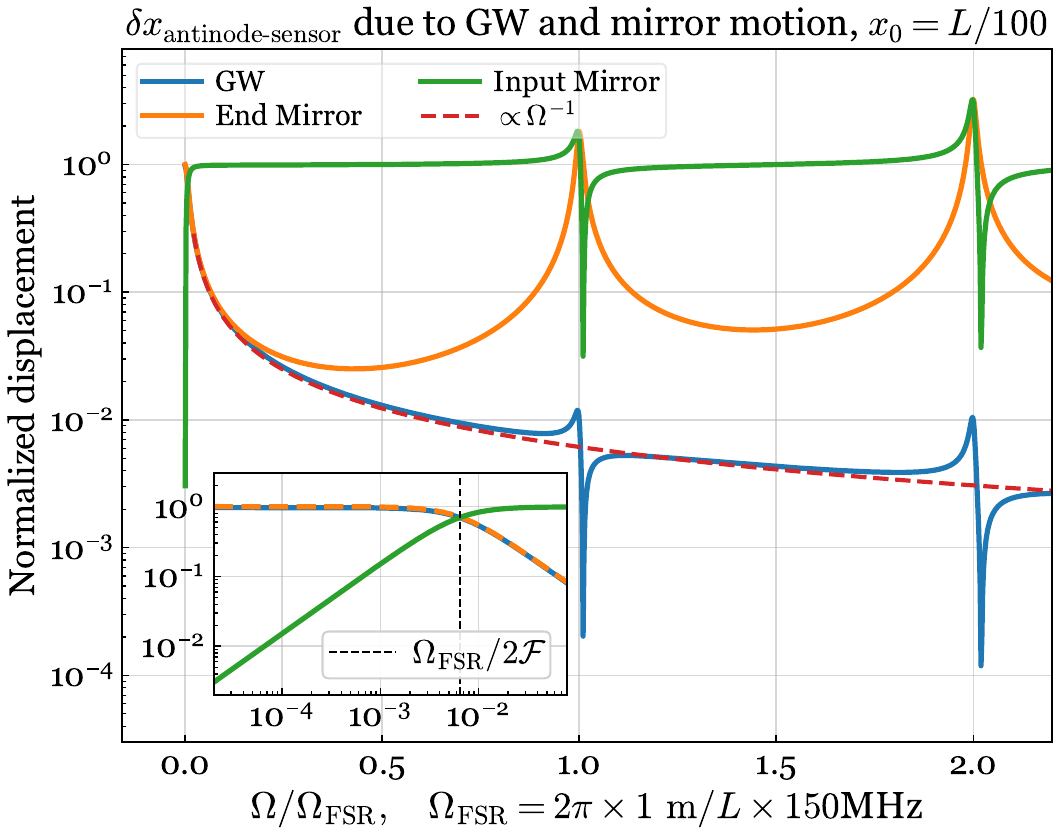}
    \caption{The frequency-dependent magnitude of the displacement of an antinode due to a gravitational wave of strain $h_0$ and mirror motion of magnitude $\delta x_\text{m}$, where $x_0=L/100$ and the cavity is on resonance. These quantities are normalized by a representative cavity length change, either $h_0L/2$ or $\delta x_\text{m}$ where appropriate. The inset shows the $\Omega\rightarrow0$ limit. The blue curve is simply a horizontal slice of Fig. ~\ref{fig:gw displacement}.}
    \label{fig:normalized antinode displacement}
\end{figure}

We can then numerically compute the resulting antinode displacement due to these perturbations. In Fig. \ref{fig:normalized antinode displacement}, we plot the magnitude of the sensor-antinode displacement $\delta x_\text{antinode}$ as a function of $\Omega$ for a GW (in the TT gauge, where the sensor is unmoved) and for mirror displacements, where $x_0$ is set to $L/100$. The mirror displacement results are normalized by $\delta x_\text{m}$, while the GW result is normalized once again by $h_0L/2$.

We observe that all three of these response functions reach local maxima at multiples of the FSR, consistent with our understanding that at these frequencies, the $\mathcal{O}(h\text{ or }\delta x_{\rm{m}})$ contribution to the total field is resonantly amplified and thus drives greater antinode displacement. We also see that the magnitude of the GW-induced antinode displacement experiences a secular decay roughly proportional to $1/\Omega$. This $1/\Omega$ behavior emerges because in full generality, the GW effect is derived from an integral along a null ray rather than an effective instantaneous position of any particular mirror.

Furthermore, the input mirror response function drops to zero when $\Omega(L-x_0)/c=\pi N$, much like the GW response function. This occurs because in this case, the sideband fields (i.e., phase variation) are generated only at the input mirror. When this condition is met, the phasors of the forward and backward propagating sideband fields will still align at the unperturbed antinode locations, having just rotated some additional integer multiple of $2\pi$ as they travel to the end mirror and back. 

In the $\Omega\rightarrow 0$ limit, the input mirror response scales with $\Omega$, but it quickly surpasses the GW and end mirror response at $\sim\Omega_\text{FSR}/2\mathcal{F}$. The fact that at most frequencies, the input mirror dominates the antinode motion is counterintuitive, especially given our argument as to why is does not move the antinode in the $\Omega\rightarrow 0$ limit. We now explain the origin of this feature.

The scale at which the transition to input mirror dominance occurs is roughly equal to the decay rate $\kappa$; when the sideband frequency surpasses $\kappa$, the consecutive reflected fields no longer interfere constructively, so the sideband field does not create a standing wave. Instead, the perturbative component takes the form of a traveling wave which is dominated by the outgoing component generated at the moving input mirror and its first reflection off the end mirror. Because the boundary condition imposed by the end mirror is the same for both the carrier and sideband fields, the misalignment angle between the phasors of the forward and backward propagating sideband fields at the \textit{unperturbed} antinode positions is roughly equal to $\Omega$ times the difference in propagation distances. As such, we expect misalignment between the sideband phasors of $\sim 2\Omega(L-x_0)/c$, and thus a shift in the antinode position by an amount $\propto (L-x_0)$.

To extract the amplitude of this proportionality, we can expand the full analytic field solution in the limit where $\kappa<\Omega\ll\Omega_\text{FSR}$. Doing so and computing the perturbed maxima of $EE^*$ gives a time-dependent antinode displacement of
\begin{multline}
    \delta x_\text{antinode}(t)\approx\delta x_\text{m}(1-x_0/L)\\\times\Bigr(\cos(\Omega t)
    -\frac{\kappa}{\Omega}\Bigr\{\sin(\Omega t)-\frac{\Omega L}{c}\cos(\Omega t)\Bigr\}\Bigr).
\end{multline}
The $(1-x_0/L)$ scaling is immediately evident, and the amplitude is $\delta x_\text{m}$ in the $x_0\rightarrow 0$ limit, matching the results of Fig. \ref{fig:normalized antinode displacement}. The term $\propto\kappa/\Omega$ captures the contribution from previous reflections, decaying as the sideband frequency continues to depart from resonance.

In accordance with this result, we have verified that when the sensor is placed at the opposite end of the cavity, the input mirror response still surpasses the GW response at $\Omega_\text{FSR}/2\mathcal{F}$ and levels off to $\delta x_\text{m}(1-x_0/L)$. With $x_0\rightarrow L$ here, it is the end mirror response which sits at a normalized value of $\sim 1$ in the $\kappa<\Omega\ll\Omega_\text{FSR}$ regime. The argument for this behavior is much like the one we just described, with the additional note that even though the carrier fields do not experience a simple reflecting boundary condition at the input mirror (due to the injected field contribution), the sideband fields still do because the input mirror is not moving in this case.

\begin{figure}
    \centering
    \includegraphics[width=\linewidth]{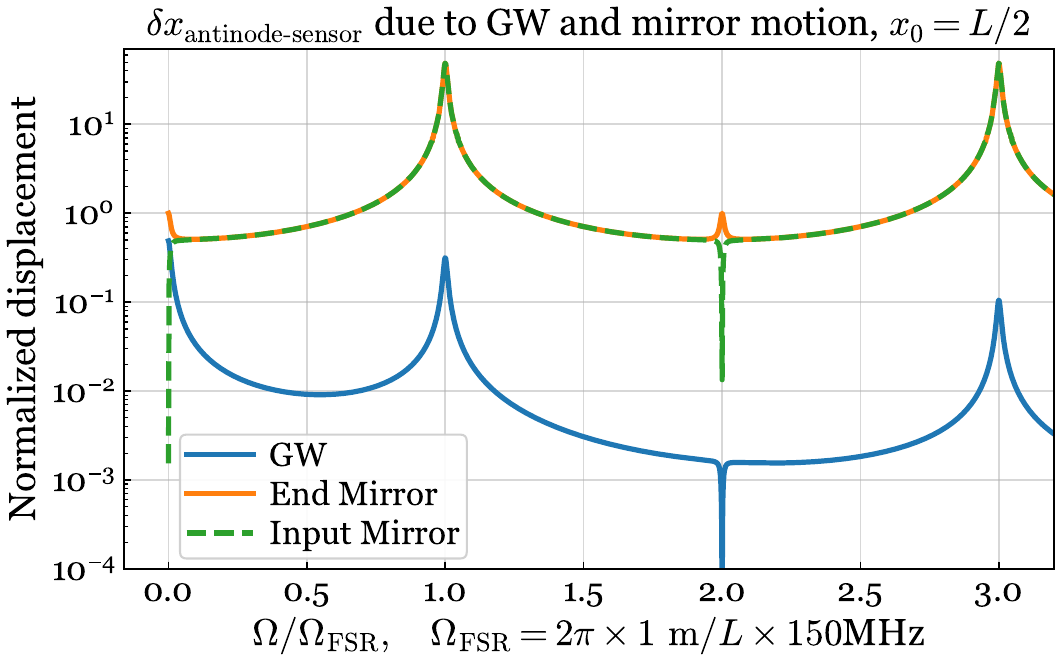}
    \caption{Identical to Fig. ~\ref{fig:normalized antinode displacement}, except we have set $x_0=L/2$ instead of $L/100$. All three response functions are maximized at odd multiples of the FSR due to resonant amplification of the sidebands, but the GW response function goes to zero at even multiples of the FSR while the mirror motion responses approach their DC limits.}
    \label{fig:normalized antinode displacement midway}
\end{figure}

As a final reference point, in Fig. \ref{fig:normalized antinode displacement midway}, we reproduce Fig. \ref{fig:normalized antinode displacement} with the exception of placing the sensor at the middle of the cavity ($x_0=L/2$) rather than near the input mirror. For frequencies $\kappa<\Omega\ll\Omega_\text{FSR}$, the normalized displacement due to motion in either mirror is 1/2, consistent with the intuition developed above. Furthermore, all three response functions are maximized at odd multiples of the FSR, where the sideband fields are resonantly amplified. At even multiples of the FSR, the GW responses goes to zero, as this coincides with the condition $\Omega(L-x_0)/c=\pi N$ being satisfied (see Sec. \ref{sec: general freq gw}).

Interestingly, at even multiples of the FSR, both mirror responses actually approach their respective DC limit. This can be explained by considering the waves which meet at $x_0$ at any given time. Since $x_0=L/2$, the forward and backward propagating waves at that point sampled the end (input) mirror position at times separated by $L/c$, that is the backward wave at $t-L/2c$ ($t-3L/2c$) and the forward wave at $t-3L/2c$ ($t-L/2c$). When $\Omega$ is an even multiple of the FSR, the phase of the mirror oscillation would have been identical at both of these retarded times. The resulting stroboscopic effect ensures that when $x_0=L/2$, the antinode displacement due to motion of each mirror behaves like its respective DC limit in the vicinity of even multiples of the FSR, matching Fig. \ref{fig:normalized antinode displacement midway}.

Earlier in this work, we argued that the displacement of the antinode offers a valid proxy for the total force on the levitated sensor, as the force from the trap beam should scale with the displacement from the minimum of the potential. This correspondence holds well at low frequencies, but more work is needed to verify whether the frequency-dependent force on the sensor continues to scale with $\delta x_\text{antinode-sensor}(t)$ at general $\Omega$, particularly when the complete field cannot be expressed as a single standing wave in the coordinate along the cavity axis (i.e. when $\Omega\sim\Omega_\text{FSR}$). We will address this question in a future work where we comprehensively compute optical forces and transfer functions for this class of system, including the effect of the sensor on the electric fields within the cavity.

\section{Implications for noise propagation and detector design}
\label{sec: noise implications}

In this section, we will briefly outline some important consequences of the asymmetry in the antinode response on how noise entering at certain ports in the experiment propagate to the detector readout channel. We will return to the perspective where $\Omega L/c\ll1$, which reasonably describes the targeted detector parameter space. 

Consider a noise source which manifests as a displacement of the input mirror surface, such as thermal noise in its coating. The motion of the input mirror surface will create optical sideband fields which propagate directly to the readout channel, but because the antinode location is insensitive to low frequency input mirror oscillations\footnote{This also assumes that the signal frequency $\Omega$ is much greater than the cavity lock loop frequency.}, the noise will not couple to the readout through motion of the levitated sensor driven by the trap beam. 

Meanwhile, an incident GW does move the antinode location, so near the resonant frequency of the optical trap, the response of the experiment to a GW should be dominated by the contribution arising from the resonantly-amplified motion of the sensor. Thus, input mirror displacement noises should be highly suppressed in units of GW strain. 

In fact, we can compute the level of suppression by examining the low-frequency regime of the mirror motion and GW response functions. For example, in the inset of Fig. \ref{fig:normalized antinode displacement}, we see that at $\Omega\sim 10^{-4}~\Omega_\text{FSR}$, which is about 15 kHz for a 1 m cavity, the input mirror only drives antinode displacements with a magnitude of $\sim10^{-2}~\delta x_\text{m}$. Therefore, in comparing a GW signal to noise in the input mirror location which causes an equivalent cavity length change, and assuming the trap resonant frequency is tuned to match $\Omega$, input mirror displacement noise at this frequency should be suppressed by a factor of $\sim100$.

This suppression does not occur for end mirror displacement noises, however, as they do drive antinode displacements and thus couple through the trap laser-driven dynamics of the sensor. These conclusions suggest that detector designs which operate at $\Omega\ll\kappa$ may benefit from targeted reduction of end mirror displacement noise through techniques such as mechanical suspension.

This asymmetry also implies that common-mode noise sources which coherently displace both the input and end mirrors by identical amounts\footnote{Seismic noise, for example, may produce these coherent displacements, assuming the cavity mirrors are mounted identically, given the tabletop size of the detector.} can effectively propagate to the readout, despite not changing the cavity length. Even though the length remains fixed, the antinodes move relative to the inertial levitated sensor, producing a driving force and thus generating resonantly-amplified motion in the vicinity of the trap frequency.

Intuitive descriptions of Fabry-Pérot-Michelson interferometers for GW detection often operate from the perspective of measuring the tiny fluctuations in the separation between the input and end mirrors. Despite also featuring Fabry-Pérot cavities, this proposed resonant detection scheme instead revolves around the separation between the sensor's equilibrium point and the end mirror. 
The input mirror instead acts here as an analogous power-recycling mirror, enabling higher resonant frequencies for the optical trap, while the levitated sensor and the end mirror form the metaphorical meterstick whose length fluctuations convey the GW strain.

For detector designs intended to operate at low frequencies (i.e., $\Omega \ll \kappa$), these noise couplings suggest that it is especially important to suppress end mirror displacement noise, such as by applying low-noise optical coatings or suspending that mirror. However, the results in Figs. \ref{fig:normalized antinode displacement} and \ref{fig:normalized antinode displacement midway} demonstrate that this coupling hierarchy can quickly reverse as $\Omega$ grows, with the relative antinode displacement due to mirror displacements greatly exceeding the GW-driven effect once $\Omega$ is comparable to the FSR. Thus, in designing detectors which will operate at frequencies beyond the decay rate, displacement noise in both mirror surfaces will need to be suppressed to a much greater extent than designs targeting low frequencies, which may prove to be a challenging task.

One additional consequence of these results is in the role of frequency noise in the trap laser. From Eq. \eqref{eq: unperturbed antinodes left}, one can show that a slow frequency deviation $\delta\omega_0$ shifts the antinode located at $x_0$ by $\delta x_\text{antinode}$, given by
\begin{equation}
    \delta x_\text{antinode}=(L-x_0)\times\delta\omega_0/\omega_0.
\end{equation}
Comparing this to the GW-induced antinode displacement, one can infer that for a single-arm detector, $\delta\omega_0/\omega_0$ would need to be reduced below $h$, which may be $\sim 10^{-22}$ or smaller \cite{SpragueAxionClouds}, to ensure the signal would not be drowned out by frequency noise. Fortunately, a two-arm Michelson setup could suppress this noise, as the antinode shift due to frequency deviations would be in the same direction in both arms, whereas a GW would shift the antinodes in each arm in opposite directions.

\section{Conclusion}
\label{sec: conclusion}
As the space of GW detector configurations continues to grow, careful computation of the effect of an incident GW on the detector observable becomes increasingly important -- such calculations can reveal key subtleties that typical approximations obscure. In this work, we have provided a fully relativistic derivation of the displacement of the antinode of a standing wave in a Fabry-Pérot cavity relative to a small levitated sensor and demonstrated gauge invariance between two of the most commonly employed gauges in GW detection analyses. This displacement serves as the starting reference for the resulting force on the levitated sensor, which drives its motion that is read out by the weak probe beam. Importantly, we have reproduced an asymmetry in the observable with respect to the sensor position which was first noticed in \cite{LSD2012}, providing the first intuitive explanation of its origin from multiple distinct gauge perspectives. Furthermore, we highlighted the implications of this asymmetry on certain noise sources -- in particular the strong suppression of input-mirror displacement noise relative to end-mirror and common-mode noise at low frequencies.

In addition to the derivation, we also extended the calculation to general values of the gravitational wave frequency, where effects due to the finite speed of light propagating in the cavity become non-negligible. These results deviate substantially from the low-frequency regime, thus serving as a crucial reference for design considerations in the development of future detectors with longer baselines.

The present analysis makes several simplifying assumptions that warrant discussion. We have assumed a specific GW orientation and polarization; the full antenna pattern and polarization dependence of the sensor response remain to be worked out and are left for future work. 
Additionally, our model assumes that the sensor can be treated like a small, sphere-like particle, more commonly used in levitated optomechanical systems \cite{ChangSpheres}, which does not significantly perturb the intracavity field. In practice, the sensors proposed for this class of detector have plate-like geometries with large cross-sections \cite{LSD2020}.
The large planar surfaces of these proposed sensors makes scattering occur primarily along the cavity axis, modifying the global field equations and boundary conditions and effectively splitting the cavity into a coupled system \cite{Laeuger2026}. 
While we expect the results presented here to remain valid under these conditions, verification will be possible once a complimentary analysis incorporating the coupled equations is complete.

This derivation and the formalism applied here are intended to complement a forthcoming paper \cite{Laeuger2026} where we outline a mathematical framework for studying the complete optomechanical response of this class of detector to gravitational waves and general noise sources, including quantum noise from both lasers, cavity response, and the modified sensor response due to the intracavity field. That analysis will also incorporate the coupled cavity effects.
Important milestones towards the completion of this experimental proposal have been recently achieved, including the successful levitation of a high aspect-ratio single material dielectric in a dual-beam optical trap \cite{LSD2022}. Detectors currently under construction will soon be in a position to directly verify the noise asymmetry result identified here, providing an important experimental test of the theoretical framework developed in this work.

\begin{acknowledgments}
A.L. is grateful for support from the Fannie and John Hertz Foundation through a Hertz Fellowship. N.A. is supported by the Noyce Foundation. A.G. and S.L. ~acknowledge support from the the W.M.~Keck Foundation. A.G. acknowledges support from NSF grants PHY-2409472 and PHY-2110524, DARPA, the John Templeton Foundation, the Gordon and Betty Moore Foundation Grant GBMF12328, DOI 10.37807/GBMF12328, the Alfred P.~Sloan Foundation under Grant No.~G-2023-21130, and the Simons Foundation. We thank Rai Weiss, Yanbei Chen, and Brian Seymour for insightful discussions throughout the progress of this work.

The data that support the findings of this article are openly available \cite{public_repo}.

\end{acknowledgments}

\bibliography{main}
\end{document}